\documentclass{elsart}
\usepackage{amssymb}
\usepackage{graphics}
\begin{document}
\begin{frontmatter}
\begin{center}
\title{Transverse voltage in zero external magnetic fields, its scaling and violation of the time reversal symmetry in MgB$_2$}

\author{P. Va\v sek}

\address
{Institute of Physics ASCR, Cukrovarnick\'a 10, 162 53 Praha
6, Czech Republic\\}

\maketitle 
\end{center}

\begin{abstract}
The longitudinal and transverse voltages (resistances)  have been measured for  MgB$_2$ in zero external magnetic fields. Samples were prepared in the form of thin film and patterned into the usual Hall bar shape. In  close vicinity of the critical temperature T$_c$  non-zero transverse resistance has been  observed. Its dependence  on the transport current  has been also studied. New scaling between transverse and longitudinal resistivities  has been observed in the form $\rho{_{xy}}\sim d\rho{_{xx}}/dT$. Several models for explanation of the observed transverse resistances and breaking of reciprocity theorem are discussed. One of the most promising explanation is based on the idea of time-reversal symmetry violation.
 
\end{abstract}
\end {frontmatter}


\section{Introduction}
Recent discovery of a new type of superconductor MgB$_2$ renewed intensive research activity into investigating a properties associated with its superconducting state. Its critical parameters as $T_c$, 
$B_c$ and $j_c$ together with simple structure and  not difficult technology processing are promising 
for practical use. It was found that  lots of physical properties are similar to those observed in high temperature (HTS) superconductors. MgB$_2$  belongs to the type -II superconductors with all the implications resulting from vortex structure in the mixed state. Anomalous Hall effect similar to the effect observed in HTS was observed as well, for example. Although numerous theoretical attempts have been made the adequate explanation of this observation was not reached.\\
It has been  suggested \cite{Hal} that in HTS the time-reversal symmetry is broken spontaneously in the absence of  an applied magnetic field. The violation of the time-reversal symmetry and of the two-dimensional reflectional symmetry (parity) is a consequence of  the fractional statistics of the quasiparticles in Laughlin's model for superconductivity \cite{Lau}. The temperature below which these symmetries are broken is expected to coincide or to be larger then the critical temperature. The violation would lead to the appearance of an antisymmetric contribution to the resistivity tensor even in the absence of an applied magnetic fields. This can be tested by four-point measurement of the resistance (van der Pauw method - four contacts on the circumference of the sample) by means of investigation of invariance of the cross resistance  under interchange of the pairs of current and voltage contacts. If there is no time-reversal symmetry violation the reciprocity relation R$_{13,24}$= R$_{24,13}$ should appear (first pair of numbers determines current contacts, last one voltage contacts).
\\
In this article we would like to show  data observed in MgB$_2$ namely the non-zero transverse voltage in the close vicinity of  critical temperature in zero external magnetic field connected with the breakdown of the reciprocity relation.
\section{Experimental}
All the measurement has been made on thin films of  MgB$_2$  of thickness about 200 nm prepared by deposition on silicon (100) single crystal substrate \cite{Plec}. Their critical temperature $T_{c0}$ is  lower then maximal critical temperature achieved on samples 
prepared on other substrates and also the width of the transition area (approx. 5K)  is  wider then in the best samples. However, these properties are desirable for our study. They enable us to investigate the effect more thoroughly to see any details in the transition region.
The sample was formed  into the shape  suitable for six-point measurement by standard photolithography . The width of active part of the sample was 100 $\mu$m. Contacts have been 
made by applying In drops in the contact area and leads then have been  soldered to this points. Disturbing voltage signals (thermoelectric etc.) have been eliminated by switching off and/or reversing  the transport current in the sample under study. The misalignment of transverse contacts was corrected  for by measuring the transverse voltage in the regime where no Hall voltage should appear, i.e. well above the transition temperature, where the sample is in the normal state.
\section{Results and discussion}
Results of the measurement in zero external magnetic field are shown in following figures.
Fig. 1 shows the resistivity transition for different transport currents through the sample. One can see a small shift of the critical temperature with increasing current density. The dependence of $T_c$  (determined as the middle point on the transition curve) on current $I$ is shown in Fig. 2. This dependence is shown there for two different current scales. One can see that the shift of $T_c$ is  proportional to $I^{2/3}$ rather then $I^{2}$, showing that heating is not appreciable and shift is connected with depairing current \cite{Ku}. 
Fig. 3 shows current dependence of transverse resistivities determined in zero external magnetic field.
In a close vicinity of the critical temperature  non-zero transverse voltage is observed.
 Its value reaches  maximum at $T_c$ and  far away from $T_c$, both above and below no such voltage has been detected , i.e. in superconducting and normal state it is zero. The value of the transverse resistivity depends on the value of the transport current. The maximum is shifted to the lower temperature    in accordance with $T_c$ and its height  decreases with increasing current. 
In Fig. 4 the comparison of the transverse resistivity peak and the derivative of the resistivity transition is shown. It is easily visible that derivative of longitudinal resistivity copies the transverse resistivity peak.

 Existence of the non-zero transverse resistance in zero magnetic field has been recently observed in different HTS materials \cite{Fra,Ja,Va,Yam}which again confirms the close link among MgB$_2$  and other HTS superconductors. In Ref. \cite{Fra} the transverse voltage was observed on thin films of YBaCuO. Similar effect has been observed recently on polycrystalline Bi$_2$Sr$_2$Ca$_2$Cu$_3$O  and YBaCuO prepared from single domain pellet \cite{Ja,Va}.  While the van der Pauw method has been used in Ref. \cite{Fra}, the classical six point method has been also used for verification of the effect \cite{Va}. \\
  Glazmann model \cite {Glaz} has been used  for the explanation of observed results  in Ref. \cite{Fra}.  In this model an isolated vortex and antivortex may be induced by transport current through a sample on the opposite sides of the sample.
Glazmann  supposes that such induced vortices and antivortices are driven to opposite sides of the sample. The vortices can meet oncoming antivortices inside the sample.These vortices and antivortices are attracted to each other and they annihilate in the end. This attraction changes movement of the vortex  in direction determined by transport current and the antivortex in the opposite direction. Both vortex and antivortex motion induces thus transversal electric field according to the Josephson relation.\\
However, in spite of Glazmann conclusion it seems to us that macroscopic mean transversal electric field must be zero for his model. In zero magnetic fields the system has vortex-antivortex symmetry.  Thus, the probability that vortex on the left meets antivortex on the right-hand side is equal to the probability of the opposite event, from whence mean transverse voltage is zero.  However, a non-zero voltage can be observed, when these induced vortices are moving under influence of any force which violates vortex-antivortex symmetry (for example guiding force \cite{Va1} etc.).\\
But vortices can be generated in the sample in the absence of the external magnetic field also by another way. In superconductors,  vortex-antivortex pairs may be excited as thermal fluctuations \cite{Jens}.
Spontaneous creation of the free vortices and antivortices has been considered in Ref. \cite{Va}as a result of thermally  activated dissociation   of vortex-antivortex pairs at Kosterlitz-Thouless temperature T$_{KT}$ \cite{KT}.
 These free vortices and antivortices can move under the influence of external electric field and cause in this way dissipation which can in principle leads to the appearance of transverse voltage. Nevertheless there should again exist  some force which will  drive vortices in the way to contribute to the non-zero transversal electric field. Moreover the Kosterlitz- Thouless transition occurs in the systems in the universality class of the two-dimensional XY model and is questionable whether  MgB$_2$   belongs to such materials. One of the approaches to the investigation of this transition is the study of current-voltage isotherms described by power law $V{\sim} I^{\alpha}$. Below of  T$_{KT}$,  $\alpha$ should decrease with increasing temperature and make an 'universal jump' from 3 to 1 at T$_{KT}$.
Here we did not observe any jump in $\alpha$. But it should be noted that this jump is  strictly defined only for the $I {\rightarrow} 0$  limit making it  difficult to detect the transition experimentally in this way .  \\
As was mentioned above, the breakdown of time-reversal symmetry should lead to the violation of the reciprocity theorem. An attempt of  an investigation of violation of reciprocity relation  has been made on YBaCuO thin film in Ref. \cite{Gi} . The authors performed their experiment in the temperature range 90 - 160 K( i.e. just above the critical temperature) on  squared sample with four contacts at the corners . According to our opinion the choice of the measuring range  is the reason why they did not see any deviations from reciprocity theorem i. e.  any anomalous "zero field Hall effect".  However, it seems  from their Fig. 3 that just near 90 K the difference R$_{13,24}$- R$_{24,13}$  can start to deviate from zero. The violation of the reciprocity theorem has been recently confirmed in Ref. \cite {Ja}. \\
One of the most interesting consequences of the broken T and P symmetry is a prediction of the existence of an intrinsic orbital magnetic moment leading to the spontaneous appearance of a macroscopic magnetization at the superconducting transition temperatures.  Such moment was really observed on epitaxial thin films of YBa$_{2}$Cu$_{3}$O  \cite{Ca}.\\
Recently Kaminski \cite{Ka} et al. reported the results of an experiment in which they found a small but significant asymmetry in the photoemission intensity of BiSrCaCuO (2212 phase) when using light of different helicities. This was interpreted as indicative of a time-reversal symmetry breaking.\\
However such explanation has been questioned by Armitage and Hu \cite{AH}. They argue that there is ample evidence for low temperature structural changes in this material and so  Kaminski et al. are incorrect to infer the existence of a time-reversal breaking state from their experiment.
   It should be noted that Laughlin's anyon mechanism for  superconductivity is formally related to  BCS state with $d_{x^2-y^2}+id_{xy}$ symmetry \cite{Ro}. As far as we know there is no indication of the existence of such state in MgB$_{2}$.
The  interesting result of this paper is shown in Fig. 4. One can see that there is a close relation between observed transverse resistance and  the derivative of longitudinal resistance. This can be written in the following 
form
\begin{equation}
\rho_{xy}(T) = A1+ B1\frac{d\rho_{xx}(T)}{dT}
\end{equation}
or
\begin{equation}
\rho_{xy}(\frac{1}{T}) = A2+ B2\frac{d\rho_{xx}(\frac{1}{T})}{d(\frac{1}{T)}} 
\end{equation}
where $A1$ or $A2$ determines the shift between the derivative and  $\rho_{xy}$ and $B1$ or $B2$ describes the scaling. One can see from Fig. 4a,b that we are not able to distinguish experimentally if Eq. 1 or Eq. 2 is more appropriate for description of the observed results.\\ The physical reason for  correlation between $\rho{_{xy}}$ and  derivative of  $\rho{_{xx}}$ is  unclear at this time and further, mainly theoretical work is needed for clarification of  this point.\\
In  conclusion, we have measured  non-zero transverse voltage in zero external magnetic field. It was also shown that there is a breaking of reciprocity theorem. The observation can be explained as a result  of violation of time-reversal symmetry. Thus our data support the idea of breaking of the time-reversal symmetry in HTS.
\section{Acknowledgements}
This work was partially supported through projects AVOZ1-010-914 and K1010104. 
 
\newpage
\section*{Figure captions}

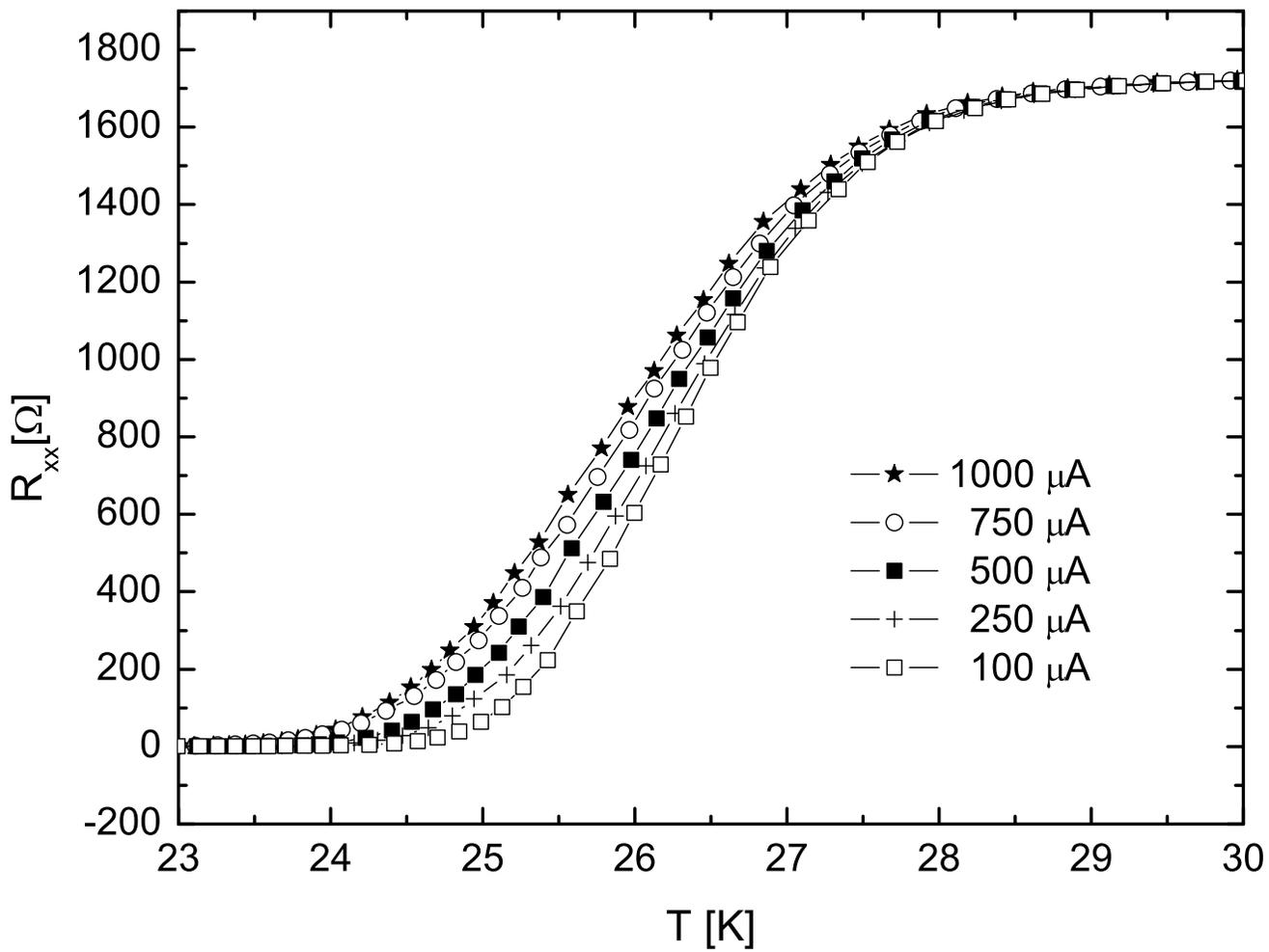
Fig. 1: Temperature dependence of the longitudinal resistance for different transport currents.

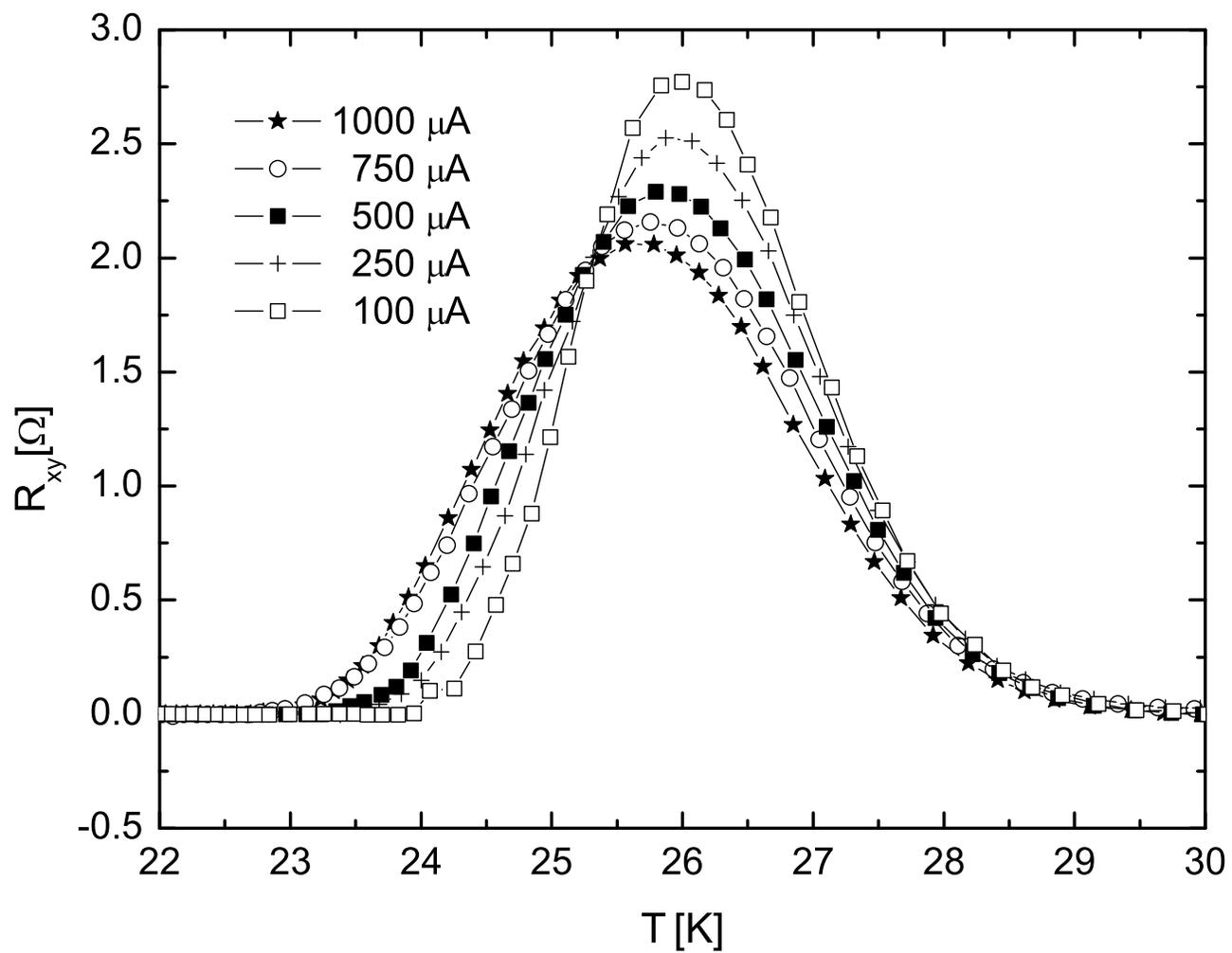
Fig. 2: Shift of transition temperature for different currents.The $I^{2/3}$ law corresponds to pair breaking, the $I^2$ is for Joule heating

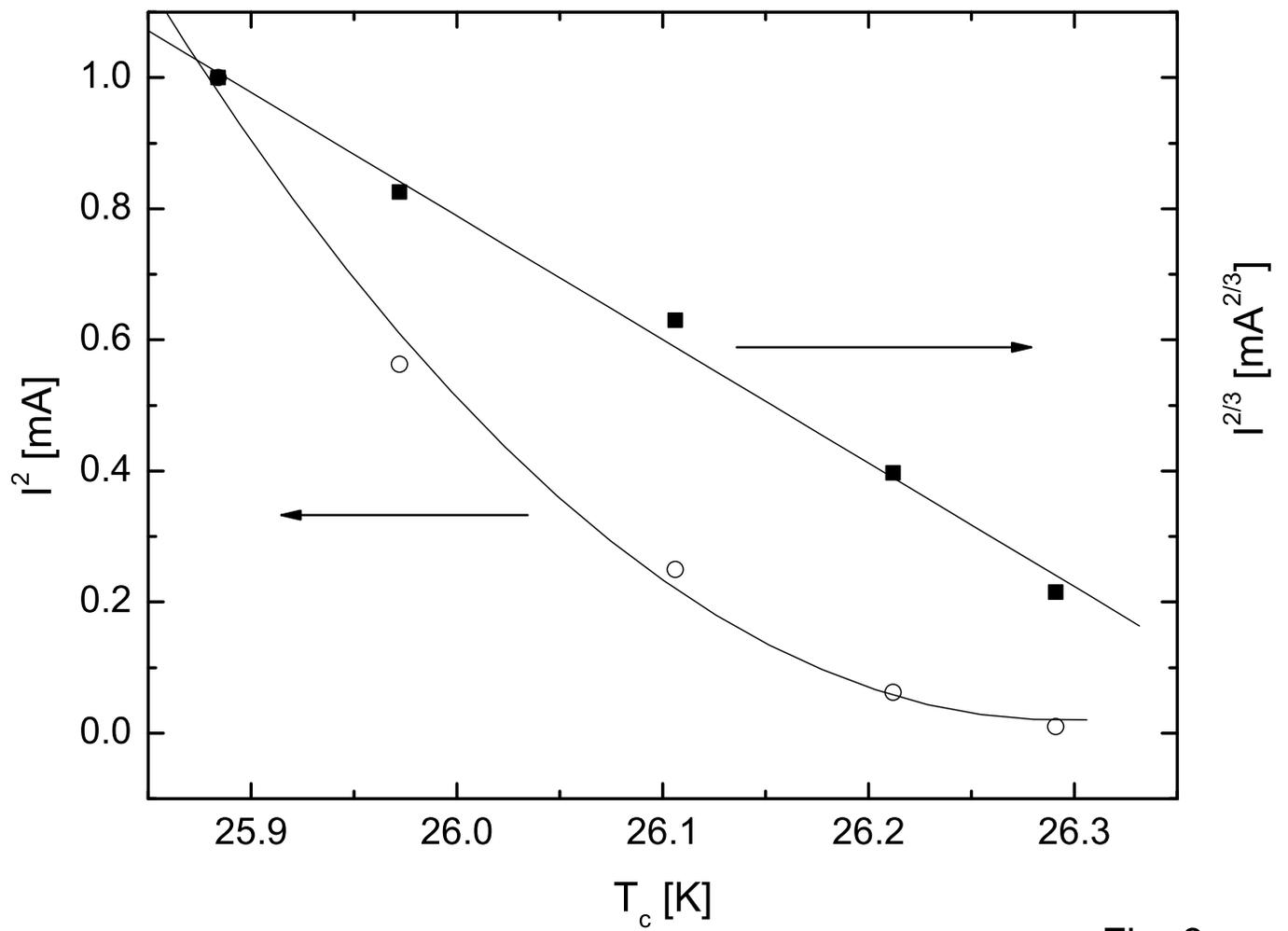
Fig. 3: Temperature dependence of the transversal resistance for different transport currents.

Fig. 4: Comparison of the  dependence of the transverse resistance ($\bullet$)  and the
 derivative of the longitudinal resistance ($\circ$) $a$ - as a function of temperature, $b$ - as a  function of inverse temperature.

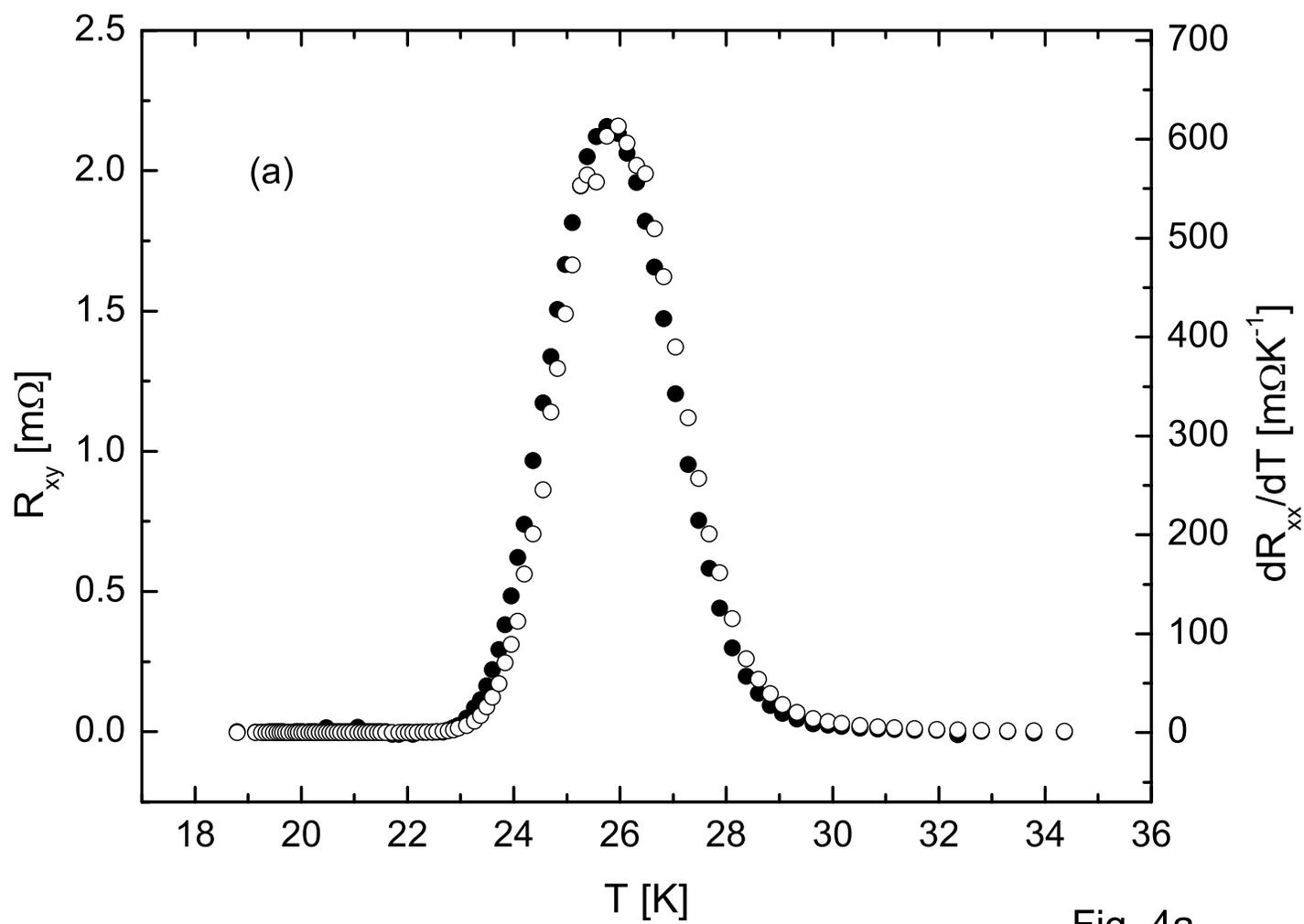
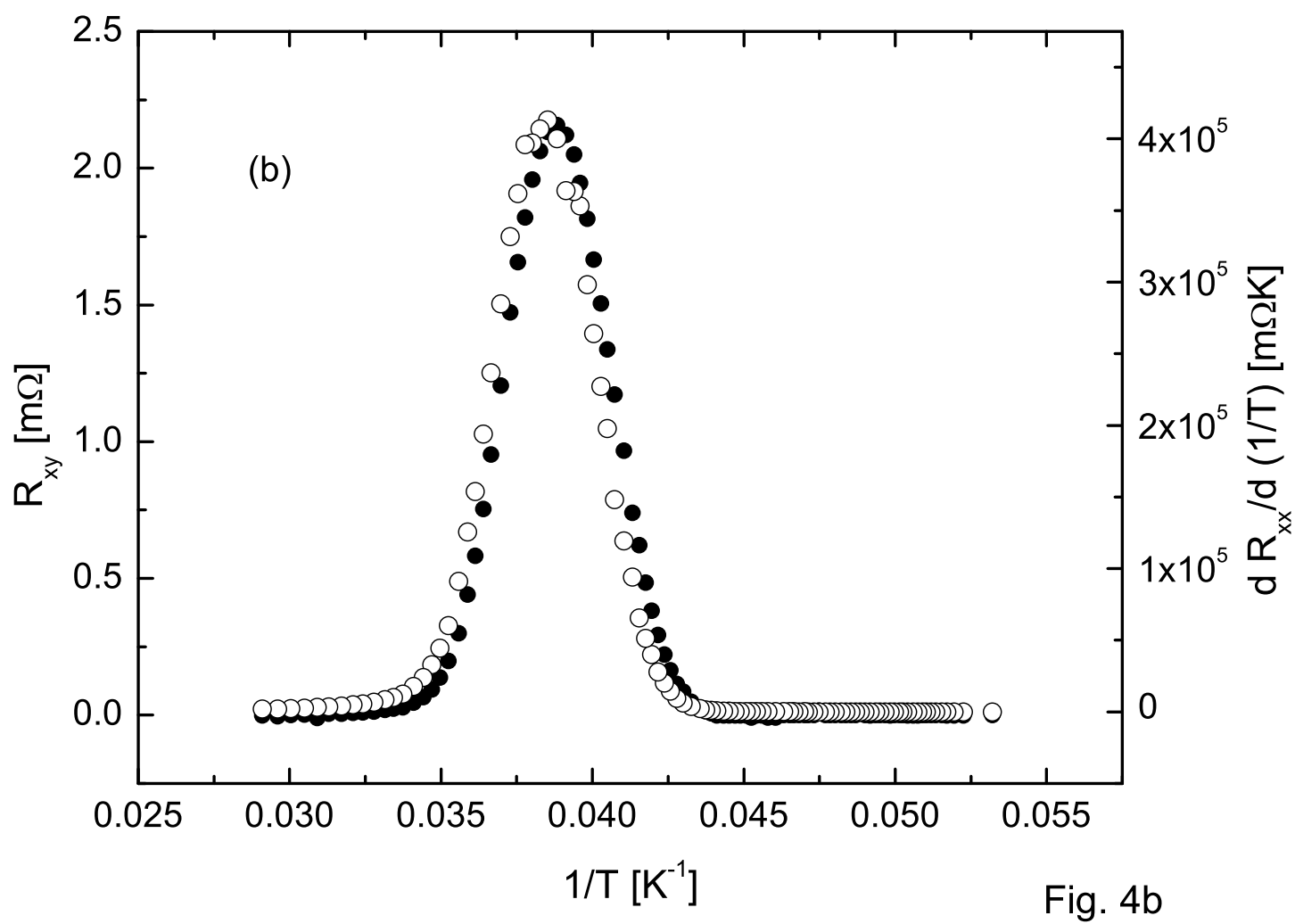
\end{document}